\documentclass[apj]{emulateapj}
\usepackage{graphics}
\usepackage{natbib}
\newcommand{\as}{$''$}

\newcommand{\nh}{$N_{\rm{H}}$}
\newcommand{\nustar}{\textit{NuSTAR}}
\newcommand{\integral}{\textit{INTEGRAL}}

\begin{document}

\title{3C\,273 with \textit{NuSTAR}: Unveiling the AGN}
\author{Kristin K. Madsen$^1$, Felix F\"{u}rst$^1$, Dominic J. Walton$^{11,1}$, Fiona A. Harrison$^1$, Krzysztof Nalewajko$^{9,14}$, David R. Ballantyne$^2$, Steve E. Boggs$^3$, Laura W. Brenneman$^4$,  Finn E. Christensen$^5$, William W. Craig$^3$, Andrew C. Fabian$^6$, Karl Forster$^1$, Brian W. Grefenstette$^1$, Matteo Guainazzi$^7$, Charles J. Hailey$^8$, Greg M. Madejski$^9$, Giorgio Matt$^{10}$, Daniel Stern$^{11}$, Roland Walter$^{13}$, William W. Zhang$^{12}$}

\affiliation{
$^1$ Cahill Center for Astronomy and Astrploophysics, California Institute of Technology, Pasadena, CA 91125, USA\\
$^2$ Center for Relativistic Astrophysics, School of Physics, Georgia Institute of Technology, Atlanta, GA 30332, USA\\
$^3$ Space Sciences Laboratory, University of California, Berkeley, CA 94720, USA\\
$^4$ Harvard-Smithsonian Center for Astrophysics, 60 Garden Street, Cambridge, MA 02138, USA 
$^5$ DTU Space, National Space Institute, Technical University of Denmark, Elektronvej 327, DK-2800 Lyngby, Denmark\\
$^6$ Institute of Astronomy, University of Cambridge, Madingley Road, Cambridge CB3 0HA, UK\\
$^7$ European Space Astronomy Centre of ESA, PO Box 78, Villanueva de la Cañada, 28691 Madrid, Spain\\
$^8$ Columbia Astrophysics Laboratory, Columbia University, New York 10027, USA\\
$^9$ Kavli Institute for Particle Astrophysics and Cosmology, SLAC National Accelerator Laboratory, Menlo Park, CA 94025, USA\\
$^{10}$ Dipartimento di Matematica e Fisica, Universita` degli Studi Roma Tre, via della Vasca Navale 84, 00146 Roma, Italy\\
$^{11}$ Jet Propulsion Laboratory, California Institute of Technology, Pasadena, CA 91109, USA\\
$^{12}$ Goddard Space Flight Center, Greenbelt, MD 20771, USA\\
$^{13}$ Geneva Observatory, University of Geneva, ch. des Maillettes 51, 1290 Versoix, Switzerland
$^{14}$ NASA Einstein Postdoctoral Fellow
}

\begin{abstract}
We present results from a 244\,ks \textit{NuSTAR} observation of 3C\,273 obtained during a cross-calibration campaign with the \textit{Chandra}, \textit{INTEGRAL}, \textit{Suzaku}, \textit{Swift}, and \textit{XMM-Newton} observatories. We show that the spectrum, when fit with a power-law model using data from all observatories except \textit{INTEGRAL} over the 1--78\,keV band, leaves significant residuals in the \textit{NuSTAR} data between 30--78\,keV. The \nustar\ 3--78\,keV spectrum is well-described by an exponentially cutoff power-law ($\Gamma = 1.646 \pm 0.006$, E$_\mathrm{cutoff} = 202_{-34}^{+51}$\,keV) with a weak reflection component from cold, dense material. There is also evidence for a weak ($EW = 23 \pm 11$~eV) neutral iron line. We interpret these features as arising from coronal emission plus reflection off an accretion disk or distant material. Beyond 80\,keV \textit{INTEGRAL} data show clear excess flux relative to an extrapolation of the AGN model fit to \nustar.  This high-energy power-law is consistent with the presence of a beamed jet, which begins to dominate over emission from the inner accretion flow at 30--40~keV. Modeling the jet locally (in the \textit{NuSTAR} + \textit{INTEGRAL} band) as a power-law, we find the coronal component is fit by $\Gamma_\mathrm{AGN} = 1.638 \pm 0.045$, $E_\mathrm{cutoff} = 47 \pm 15$\,keV, and jet photon index by $\Gamma_\mathrm{jet} = 1.05 \pm 0.4$. We also consider \textit{Fermi}/LAT observations of 3C\,273 and here the broad-band spectrum of the jet can be described by a log-parabolic model, peaking at $\sim 2$\,MeV. Finally, we investigate the spectral variability in the \textit{NuSTAR} band and find an inverse correlation between flux and $\Gamma$. 
\end{abstract}


\section{Introduction}
At a redshift of $z = 0.158$ \citep{Schmidt1963}, 3C\,273 is the nearest high luminosity quasar and has been extensively studied at all wavelengths since its discovery in 1963 \citep[for a review, see ][]{courvoisier1998}. It is radio-loud and highly variable across nearly all energies \citep{Soldi2008}, with a jet showing apparent superluminal motion. 

At radio to millimeter and at $\gamma$-ray energies, flares from the relativistic jet dominate the variability of 3C\,273 \citep{Turler2000,Abdo2010}. In the optical-UV band there is a bright excess (blue bump)  which is spectrally complicated and suggestive of two independently varying components: a more rapidly varying component originating from thermal reprocessing from an accretion disk, and a slower component that could be synchrotron emission unrelated to the radio-mm jet \citep{Paltani1998}. As observed in many other AGN, there is a soft-excess in the low-energy  ($<2$\,keV) X-ray band possibly due to Comptonized UV photons \citep{Page2004}.    Correlations between the UV and low-energy X-rays \citep{Walter1992,Page2004} have been noted in the past, supporting this interpretation. However, recent observations have failed to detect correlated variability \citep{Chernyakova2007,Soldi2008}, leaving the interpretation of the optical/UV excess uncertain. 

There is evidence of an intermittently weak iron line in the X-ray spectrum, which appears to be broad ($\sigma \sim 0.6$\,keV, EW $\sim$ 20--60\,eV), occasionally neutral \citep{Turner1990,Page2004,Grandi2004} and sometimes ionized \citep{Yaqoob2000,Kataoka2002}. The line is too faint to reliably trace its variability as a function of flux with current instrumentation. 

Above  2\,keV and up to MeV energies, previous observations report a hard power-law spectrum, as is common for jet-dominated AGN (blazars). Over the 30 years that 3C\,273 has been reliably monitored, there appears to be a long term spectral evolution underlying the short term variations. The source was in its softest observed state in June 2003 (photon index $\Gamma \sim 1.82 \pm 0.01$), a value of $\Delta\Gamma \sim$ 0.3--0.4 above what was measured in the 1980's ($\Gamma \sim 1.5$), and since then the source has hardened again to a value of $\Gamma \sim 1.6 - 1.7$ \citep{Chernyakova2007}. 

Finally, 3C\,273 is a strong $\gamma$-ray emitter; it has been routinely detected by \textit{Fermi}, showing several interesting $\gamma$-ray flares \citep{Abdo2010}. Broad-band observations including \textit{Fermi} and \textit{INTEGRAL} reported by \citet{Esposito2015} imply that the broad-band spectrum of the jet emission is not a single power-law, but instead, it can be described by a log-parabola model peaking around 3--8\,MeV.

Because of the presence of the soft excess and the iron line, it has been postulated that the X-ray emission arises from a mix of thermal and non-thermal processes: an AGN component from the inner accretion flow (i.e. emission from the corona possibly accompanied by disk reflection) and a jet component. \citet{Grandi2004} fitted the broadband spectrum from \textit{BeppoSAX} with a cold reflector irradiated by an isotropic coronal X-ray source for the AGN component \citep{Magdziarz1995} and a power-law for the jet component. They were able to obtain reasonable fits to a restricted model where the coronal continuum photon index and cutoff energy were held fixed. A search for the disk reflection hump at $\sim$30\,keV was presented in \citet{Chernyakova2007} using \textit{XMM-Newton} and \textit{INTEGRAL} data. They estimated that about 20\% reflection could be allowed, but no clear results could be obtained because of the non-overlapping energy bands between the instruments and the gap from 10--20\,keV, making cross-calibration uncertain. 

The two component scenario is supported by variability studies made by \citet{Soldi2008}, who found that the amplitude of the variations increases steeply above 20\,keV with no convincing correlation between the variations above and below 20\,keV. As of yet no clear signature above 10\,keV of a disk related component in the form of a reflection feature and/or up-scattering of the thermal disk photons by a corona has been found. We show that this is no longer the case and present here overlapping contemporaneous observations of 3C\,273 obtained by the six observatories: \textit{Chandra}, \textit{INTEGRAL}, \textit{NuSTAR}, \textit{Suzaku}, \textit{Swift}, and \textit{XMM-Newton}. We place particular emphasis on the \textit{NuSTAR} observation, which was about 6 times longer than the rest and for the first time allows a study of the 3--78\,keV region without cross-calibration concerns. 

%In \S \ref{bb} we study the broadband spectrum in the overlapping window between observatories. In \S \ref{ironcont} we focus on \textit{NuSTAR} alone and investigate the iron-line and continuum of the entire observation, while in \S \ref{variable} we study the variability of the source.

\section{Data Analysis}
\begin{figure}
\includegraphics[width=0.45\textwidth]{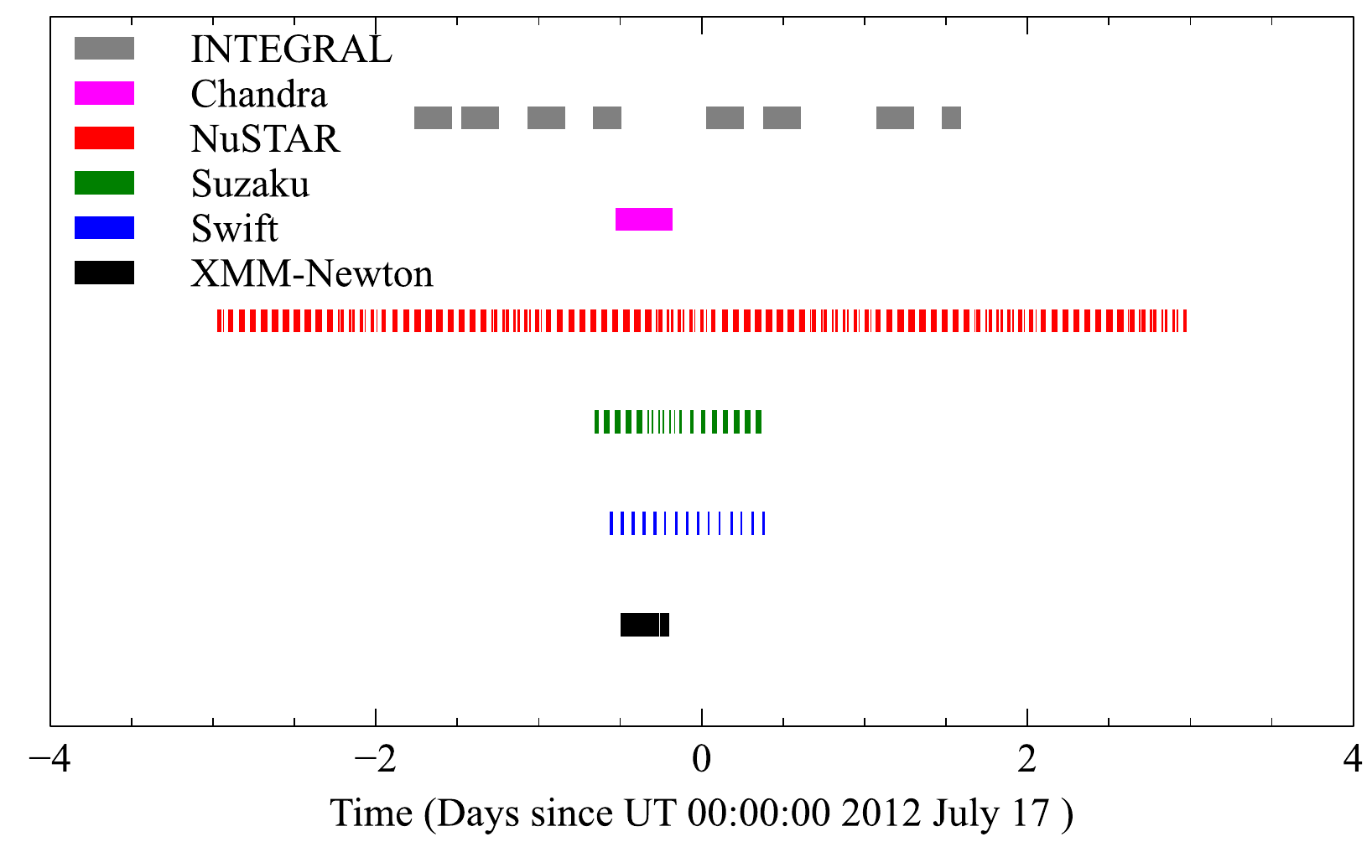}
\caption{Length and coverage of the six overlapping 3C\,273 observations. We do not enforce strict simultaneity between the low earth orbit observatories \textit{NuSTAR}, \textit{Suzaku}, \textit{Swift}, and extract the full range for all observatories - except for \textit{INTEGRAL} - between the \textit{Suzaku} start and stop times as indicated by the shaded region in Figure \ref{lightcurve}.}
\label{coverage}
\end{figure}

On UT 2012 July 17, the six observatories \textit{Chandra} \citep{Weisskopf2002}, \textit{INTEGRAL} \citep{Winkler2003}, \textit{NuSTAR} \citep{Harrison2013}, \textit{Suzaku} \citep{Mitsuda2007}, \textit{Swift} \citep{Gehrels2004}, and \textit{XMM-Newton} \citep{Jansen2001} observed 3C\,273 as part of a cross-calibration campaign organized by the International Astronomical Consortium for High Energy Calibration (IACHEC\footnote{http://web.mit.edu/iachec/}). Figure \ref{coverage} shows the overlapping GTI times and length of each observation and Table \ref{obsid} lists the observation ID's and exposure times for the respective observatories. 

For \textit{Chandra} we used CIAO 4.6.1 and CALDB 4.6.1.1. The data were taken with gratings configuration ACIS+HETG and reprocessed using the CIAO \texttt{chandra\_repro} reprocessing script. We combined orders 1--3 for the HEG and MEG arm separately, and binned the data at 30 counts. 

For \textit{INTEGRAL} we used the standard OSA 10.0 data reduction package. We only used data from IBIS/ISGRI \citep{Ubertini2003, Lebrun2003} as data from JEM-X and SPI did not allow to constrain the spectral shape significantly. We extracted the IBIS data on a science window (ScW) by ScW basis using our own scripts, which are based on the IBIS Analysis User Manual procedure. We combined the spectra using \texttt{spe\_pick} according to observation date.

For \textit{NuSTAR} we used HEAsoft 6.15.1 and CALDB 20131223. The data were processed with all standard settings and source counts extracted from a 30\as\ radius circular region. Background was taken from the same detector. 

For \textit{Suzaku} we used CALDB: HXD (20110913), XIS (20140203) and XRT (20110630). The observation was taken in 1/4 window mode, and we used 100 arcsec radius circular regions for the FI detectors (XIS0,3) and a 140 arcsec region for the BI detector (XIS1), such that the regions were centered on the source, but were restricted to the operational portions of the detectors.

For \textit{Swift} we used HEAsoft 6.15.1 and XRT CALDB 2014-02-04. The data were taken in `PHOTON' mode and were reduced using \texttt{xrtpipeline}. Spectra were extracted from an annulus region, inner radius 5\as\ and outer radius 30\as\ to correct for pileup. The two observations were combined and the spectra binned at 50 counts.

For \textit{XMM-Newton} we used SAS v. 13.5.0 with CALDB 2014-01-31. The data were taken in `Small Window' mode, and to corrected for pileup in the MOS we excised counts form an annulus region with inner radius of 15\as\ and outer radius of 45\as. For the PN we extracted from a circular region of radius 45\as.

We analyzed the high-energy $\gamma$-ray data from the Fermi Large Area Telescope (LAT), using the software package ScienceTools v10r0p5 and the latest calibration standard Pass 8. Since 3C\,273 was found in a relatively low $\gamma$-ray state, this required integrating the $\gamma$-ray flux over a timescale of 50 days (MJD 56100-56150), which is significantly longer than the 6-day \textit{NuSTAR} campaign. We used the instrument response function P8R2\_SOURCE\_V6 (front and back), including galactic diffuse emission model gll\_iem\_v06, isotropic background model iso\_P8R2\_SOURCE\_V6\_v06, and background point sources within 15 deg from 3C\,273 taken the 2FGL catalog. Events were extracted from a region of interest within 10 deg from 3C\,273, we applied standard selection cuts for the SAA avoidance, and the zenith angle cut z $<$ 100\,deg.

\begin{table*}
\centering
\caption{Observation Log}
\begin{tabular}{l||c|c|c}
\hline
Instrument & Start Time & OBSID & Exposure (ks) \\
\hline
\textit{Chandra} & 2012 Jul 16 UT 11:04:29 & 14455 & 30.0\\
\textit{NuSTAR} & 2012 Jul 14 UT 00:06:07 & 10002020001 & 244.0\\
\textit{Swift} & 2012 Jul 16 UT 10:24:59 & 00050900019	& 13.0\\
\textit{Swift} & 2012 Jul 17 UT 00:50:59 & 00050900020	& 6.9 \\
\textit{Suzaku} & 2012 Jul 16 UT 08:08:54 & 107013010 & 39.8 \\
\textit{XMM-Newton} & 2012 Jul 16 UT 11:59:23 & 0414191001 & 38.9\\
\hline
\textit{INTEGRAL} & 2011 Dec 05 UT 02:27:04 & 1116\footnote{Revolutions.}--1128 (321 ScW) & 689.5 \\
\textit{INTEGRAL} & 2012 Jun 08 UT 04:08:54 & 1178, 1180--1183, 1188 (66 ScW) & 131.6 \\
\textit{INTEGRAL} & 2012 Jul 14 UT 22:18:59 & 1191--1192 (99 ScW) & 68.4 \\
\hline
Instrument & Start Time & Stop Time & Exposure (days) \\
\hline
\textit{Fermi} & 2012 Jun 22 UT 00:00:00 & 2012 Aug 11 UT 00:00:00 & 50 \\
\hline
\end{tabular}
\label{obsid}
\end{table*}

\begin{figure}
\includegraphics[width=0.45\textwidth]{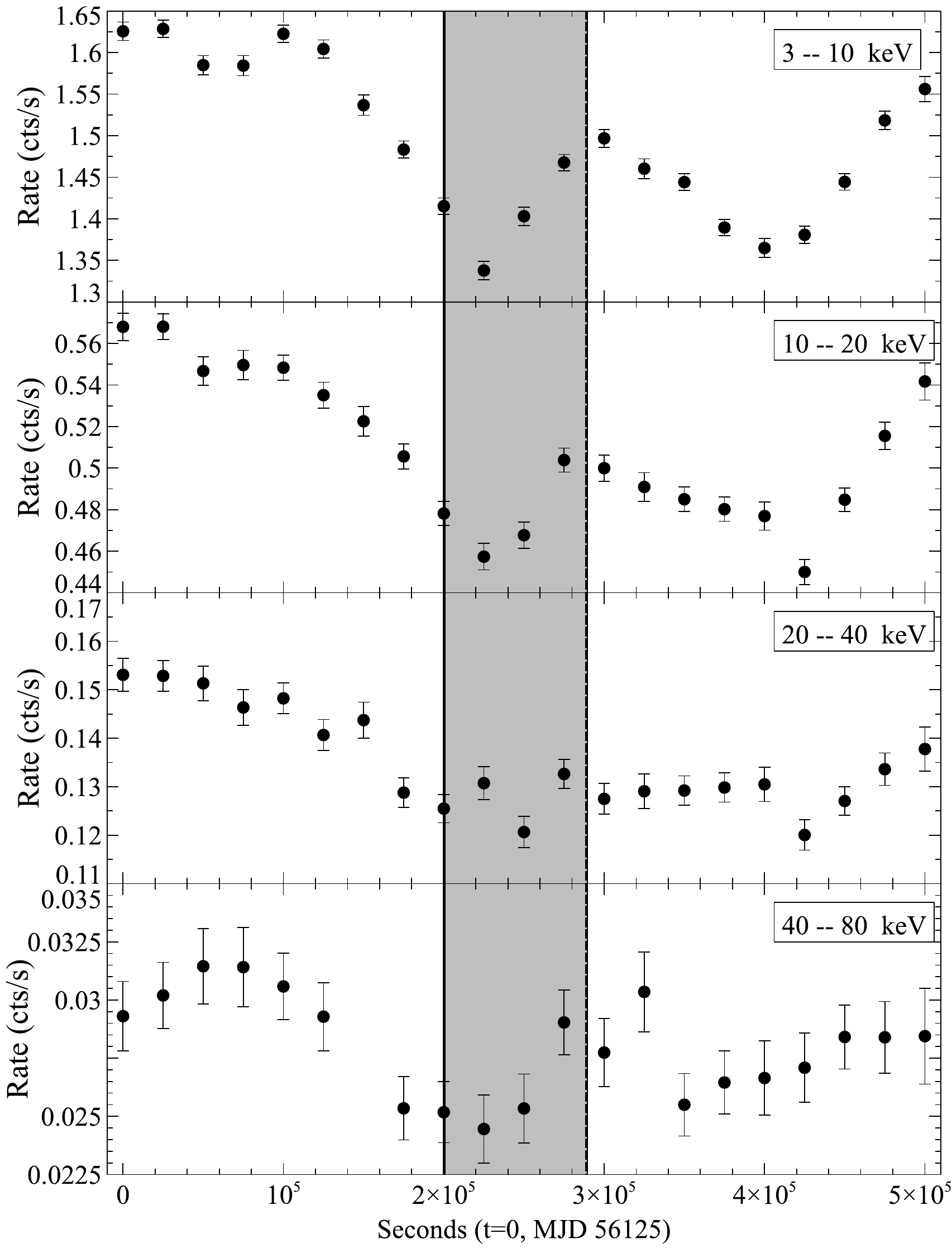}
\caption{Light-curve in different energy bands for \nustar. Bin width is 25ks and the light-curve has been live-time and vignetting corrected. The shaded region marks the start and stop of the \textit{Suzaku} observation.}
\label{lightcurve}
\end{figure}

\section{Spectral Fitting}

For all the data analysis we use the XSPEC version 12.8.2 analysis software \citep{Arnaud1996}, and present 90\% confidence limits unless otherwise stated.

\subsection{Broad-band Fitting}\label{bb}

To investigate the overall broadband spectrum we fit data to all observatories except \integral\ (to be discussed in the next section) for an interval with reasonable temporal overlap. Figure~\ref{coverage} shows the observation times for the various missions. We chose the interval spanning the \textit{Suzaku} observation for extraction of data since there is reasonable overlap of part of this interval with all observatories except \integral. Due to the relative phasing of the South Atlantic Anomaly (SAA) passages and occultation periods among the low Earth orbit observatories (\textit{NuSTAR}, \textit{Suzaku}, and \textit{Swift}), we do not enforce strict simultaneity within this window. The \textit{NuSTAR} lightcurve in Figure \ref{lightcurve} (binned at 25\,ks, live-time and vignetting corrected) shows that during the \textit{Suzaku} window, marked by the shaded region, the flux changed only gradually by $\sim$10\%. We therefore do not expect sudden flux changes to have occurred between occultation/SAA periods. We extracted the full time range for all instruments except \textit{NuSTAR} where we truncated the observation by applying the \textit{Suzaku} start and stop times. Because the majority of the \textit{INTEGRAL} observation lies outside the near simultaneous window, where the flux is significantly different, we excluded \textit{INTEGRAL} from this broadband fit. 

We fitted an absorbed power-law model using Wilms abundances \citep{Wilms2000} and Verner cross-sections \citep{Verner1996}, freezing the hydrogen column to the Galactic value of 1.79$\times 10^{20}$ cm$^{-2}$ \citep{Dickey1990}. We keep this value fixed in the remainder of the fits. To avoid complications with the soft excess, we limited our fitting range between 1--78\,keV, though we note that during these observations the soft excess appears to have been very modest. We allowed all normalizations to float; the relative cross-normalization terms and the relative slope errors between instruments are discussed in detail in Madsen (in prep) by the IACHEC consortium. 

\begin{figure}
\includegraphics[width=0.45\textwidth]{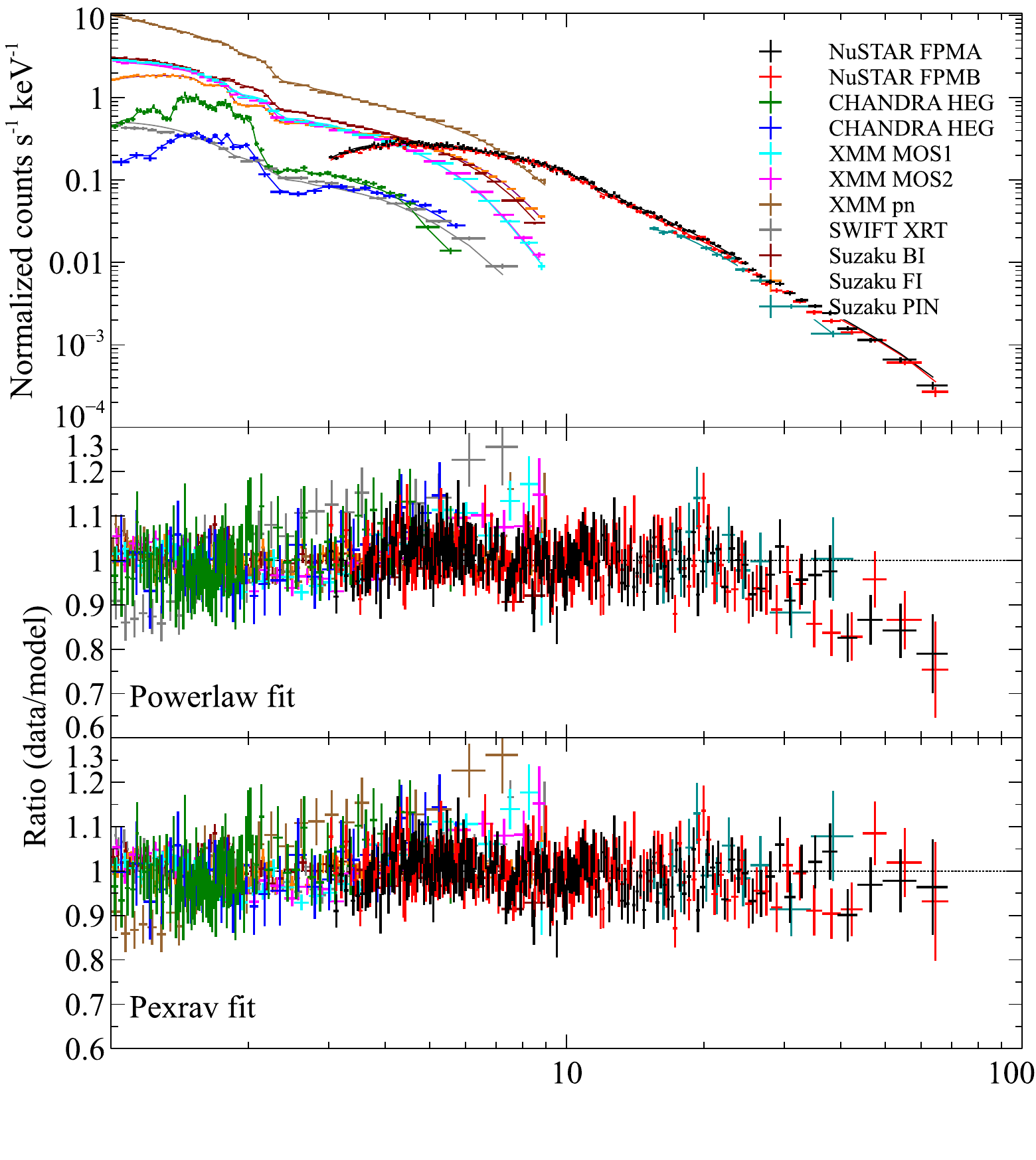}
\caption{\textbf{Top panel}: Simultaneous broadband absorbed power-law fit between 1--78\,keV to all observatories except \textit{INTEGRAL}. Cross-normalization constants are allowed to float and the column is fixed to the Galactic value of \nh=1.79$\times 10^{20}$ cm$^{-2}$. The best fit photon index is $\Gamma = 1.647 \pm 0.003$. \textbf{Bottom panels}: Ratio of a power-law (middle panel) and the \texttt{pexrav} model (bottom) shown for \textit{NuSTAR} alone. \texttt{pexrav} fit: $\Gamma = 1.646 \pm 0.006$, E$_\mathrm{cutoff} = 202_{-34}^{+51}$\,keV, and relative reflection of $R = 0.15 \pm 0.05$.}
\label{broadband}
\end{figure}

\begin{figure}
\includegraphics[width=0.45\textwidth]{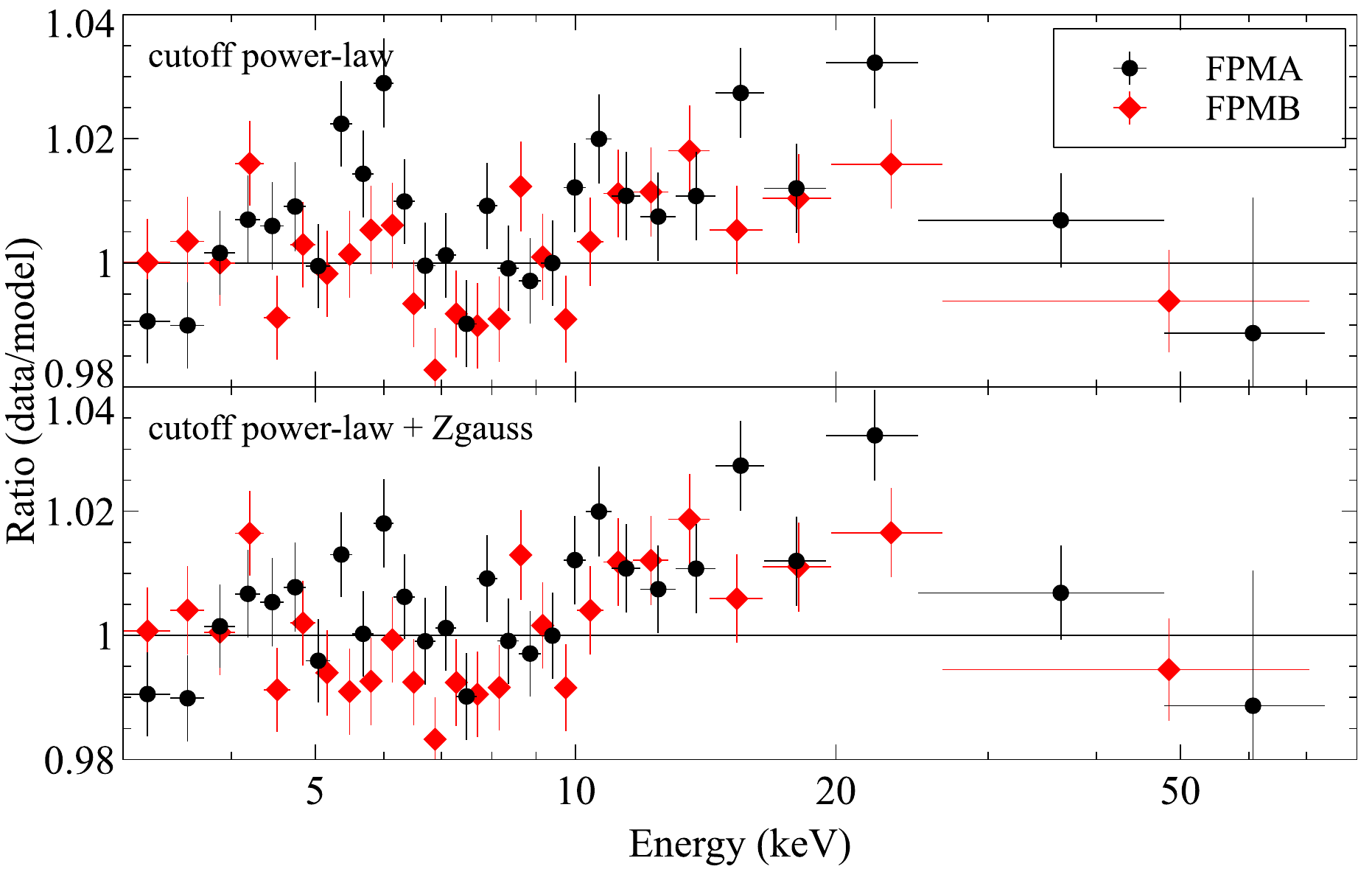}
\caption{Ratio of \textit{NuSTAR} data alone to a cutoff power-law model ignoring the energy ranges 5--8\,keV and 10--50\,keV. The \textit{NuSTAR} calibration errors are at the $\pm 1 \%$ level, which accounts for the spectrum's scatter below 5\,keV. The iron line and Compton reflection hump are very modest but systematically different from the known calibration errors.}
\label{cutoffratio}
\end{figure}

Figure \ref{broadband} shows the results of the best fit power-law: $\Gamma = 1.647 \pm 0.003$. Formally the fit is good, with $\chi^2_\mathrm{red}=1.018$ (7815 dof). However, visual inspection of the high energy residuals above 20\,keV in \nustar\ shows obvious systematic deviations from a power-law. This good formal fit is due to the very large number of bins between 2--8\,keV. Replacing the power-law with a cutoff power-law, \texttt{cutoffpwrlw} (XSPEC), reduces the residuals, and we find $\Gamma = 1.624 \pm 0.006$, $E_\mathrm{cutoff} = 291_{-55}^{+90}$\,keV ($\chi^2_\mathrm{red}=1.013$). To examine any further features we fitted a cutoff power-law to \nustar\ alone, ignoring the energy ranges 5--8\,keV and 10--50\,keV and plot the \textit{NuSTAR} ratio of the model to data alone in Figure \ref{cutoffratio}. The shape of these residuals is reminiscent of reflection of the primary continuum off dense material, with an iron line and a Compton hump. The \textit{NuSTAR} calibration uncertainties are $\pm 1\%$ up to 10\,keV, $\pm 2\%$ up to 40\,keV, and 5--10\% above \citep{Madsen2015}. While the excess is at a level only somewhat greater than the calibration uncertainties, the systematic shape is not characteristic of calibration errors, and we associate it with a very weak reflection component.

We then fitted all the instruments with the combination of a primary cutoff power-law plus reflection from a plane parallel slab of neutral material (the \texttt{pexrav} model in XSPEC; see \citep{Magdziarz1995}).   The jet angle and hence disk orientation has been estimated from radio observations to be $\sim 10^\circ$ \citep{Abraham1999}, though fits to the UV-continuum and iron line suggest it could be as high as $\sim 60^\circ$ \citep{Yaqoob2000}.  Assuming the material to be associated with the accretion disk, we tested the fits for both angles, but since we found no appreciable effect on the fit, we fixed the value at 35$^\circ$. We used solar abundances and the best fit is achieved for $\Gamma = 1.646 \pm 0.006$, E$_\mathrm{cutoff} = 202_{-34}^{+51}$\,keV, with a weak relative reflection component with ratio of direct to reflected flux of $R=0.15 \pm 0.05$ ($\chi^2_\mathrm{red}=1.006$). We show the ratio of model to data of this fit in Figure \ref{broadband}. A Monte-Carlo simulation shows the likelihood of a reflection of this order to be in excess of 99.9\%.

\subsection{Fitting \textit{NuSTAR} and \textit{INTEGRAL}}\label{integralsection}
\begin{figure}
\includegraphics[width=0.45\textwidth]{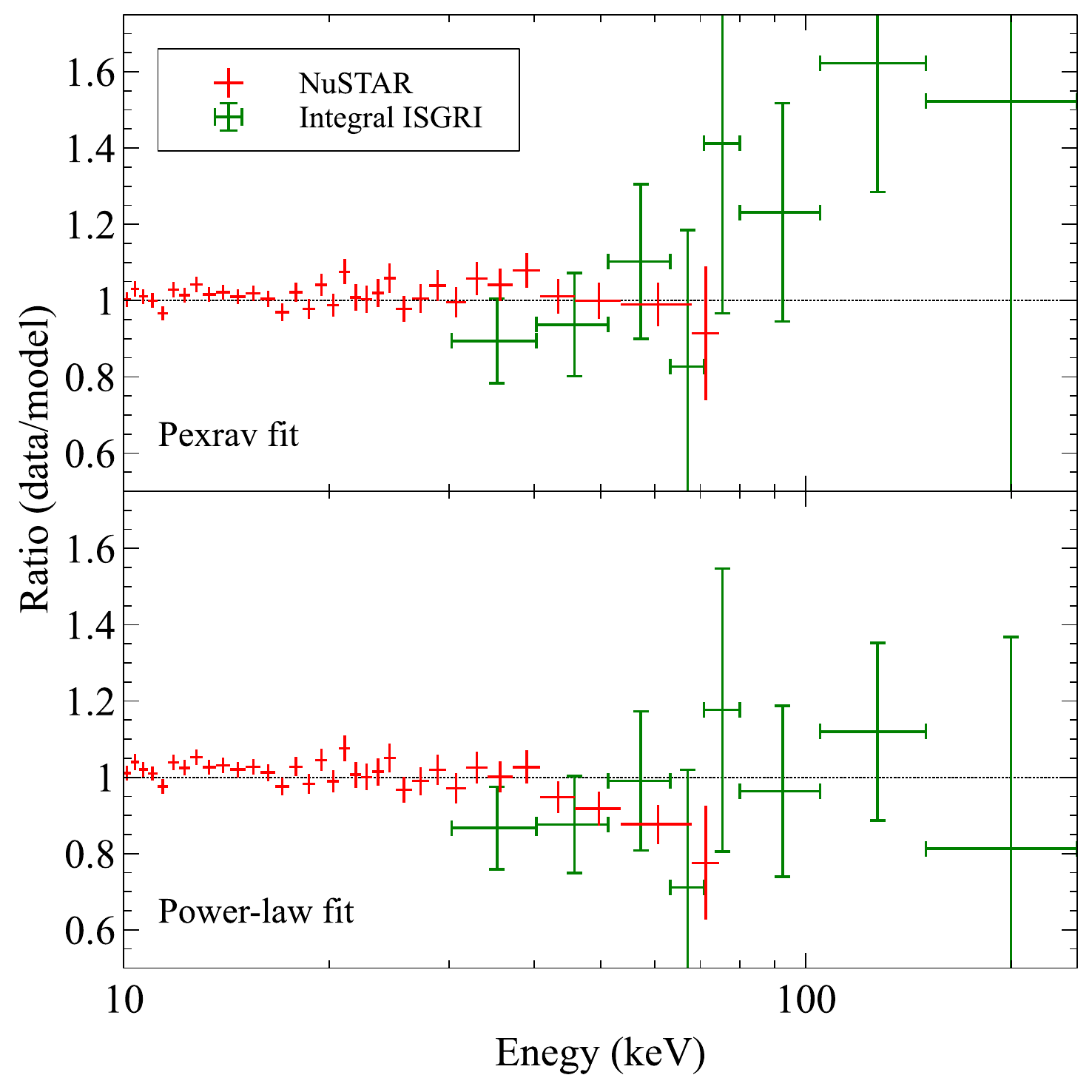}
\caption{Residuals of the data to model of the fits between \nustar\ and \integral\ for a \texttt{pexrav} (top) and \texttt{pegpwrlw} (bottom) fit. Only \textit{NuSTAR} FPMA is shown for ease of viewing.}
\label{badfitintegral}
\end{figure}

We investigated joint \textit{NuSTAR} and \textit{INTEGRAL} observations over a broader time range, dispensing with the low energy-instruments. We applied the \textit{INTEGRAL} GTI's from the July observation to the \textit{NuSTAR} data set. Fitting the data with a \texttt{pexrav} alone results in \textit{INTEGRAL} diverging above 80\,keV. A power-law provides a reasonable broad band fit, but with strong \textit{NuSTAR} residuals between 20--78\,keV as shown in Figure \ref{badfitintegral}. Following the idea that the X-ray spectrum is in reality a superposition of an AGN component (i.e. coronal emission plus disk reflection) with a jet component, we fitted the data with the \texttt{pexrav} model for the AGN component and a flux pegged power-law, \texttt{pegpwrlw} (XSPEC), for the jet.

\begin{figure}
\includegraphics[width=0.45\textwidth]{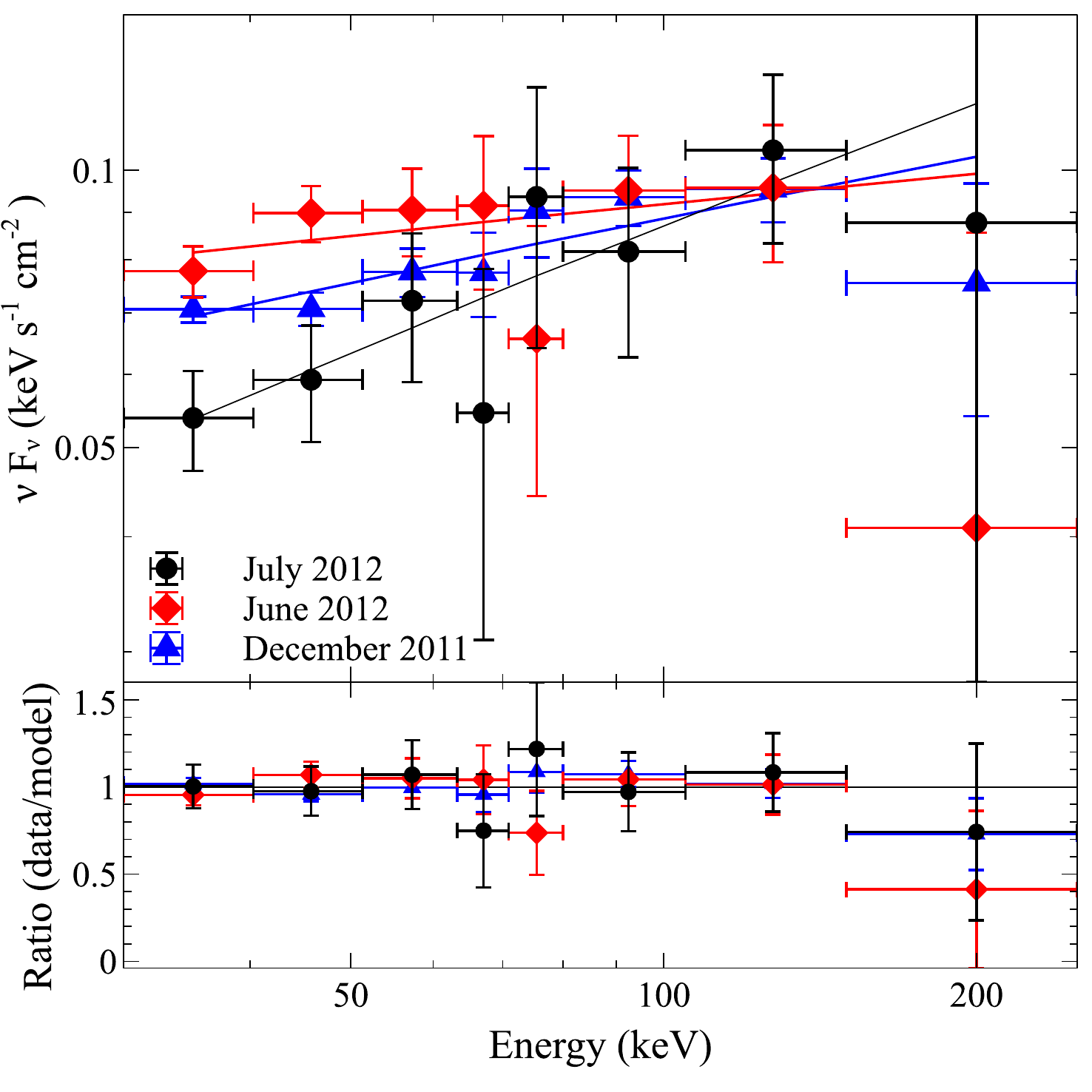}
\caption{Power-law fits to \textit{INTEGRAL} data. Flux differences decrease with increasing energy and appear to converge around $\sim 120$\,keV. Fit parameters are recorded in Table \ref{tableIntegral}.}
\label{integral}
\end{figure}

The two models prove to be degenerate. To partially break this degeneracy we investigated three \textit{INTEGRAL} data sets taken in December 2011, June 2012, and July 2012. We fitted the spectra with \texttt{pegpwrlw} from 30--250\,keV. The individual fits are summarized in Table \ref{tableIntegral} with the spectral energy distributions shown in Figure \ref{integral}. The spectral slope changes with time, but the flux difference decreases with increasing energy and converges at $\sim 120$\,keV. We also draw attention to the dip in the June and July observations between 60--80 \,keV; the error bars are large, but this could be interpreted as the AGN turn-over seen by \textit{NuSTAR}. 
%Finally in all data sets the last high energy bin is systematically low likely due to issues with background subtraction.

Based on the fact that for different slopes of the spectrum in \textit{INTEGRAL} the flux between 80-150\,keV remained almost constant, we decided to freeze the flux of the jet component between 80--150\,keV to $92\times10^{-12}$ erg cm$^{-2}$ s$^{-1}$ as obtained from the July observation. Additionally, we freeze the relative reflection to R=0.15 to further limit the degeneracy, but allowed all other parameters to vary. 

\begin{figure*}
\includegraphics[width=0.95\textwidth]{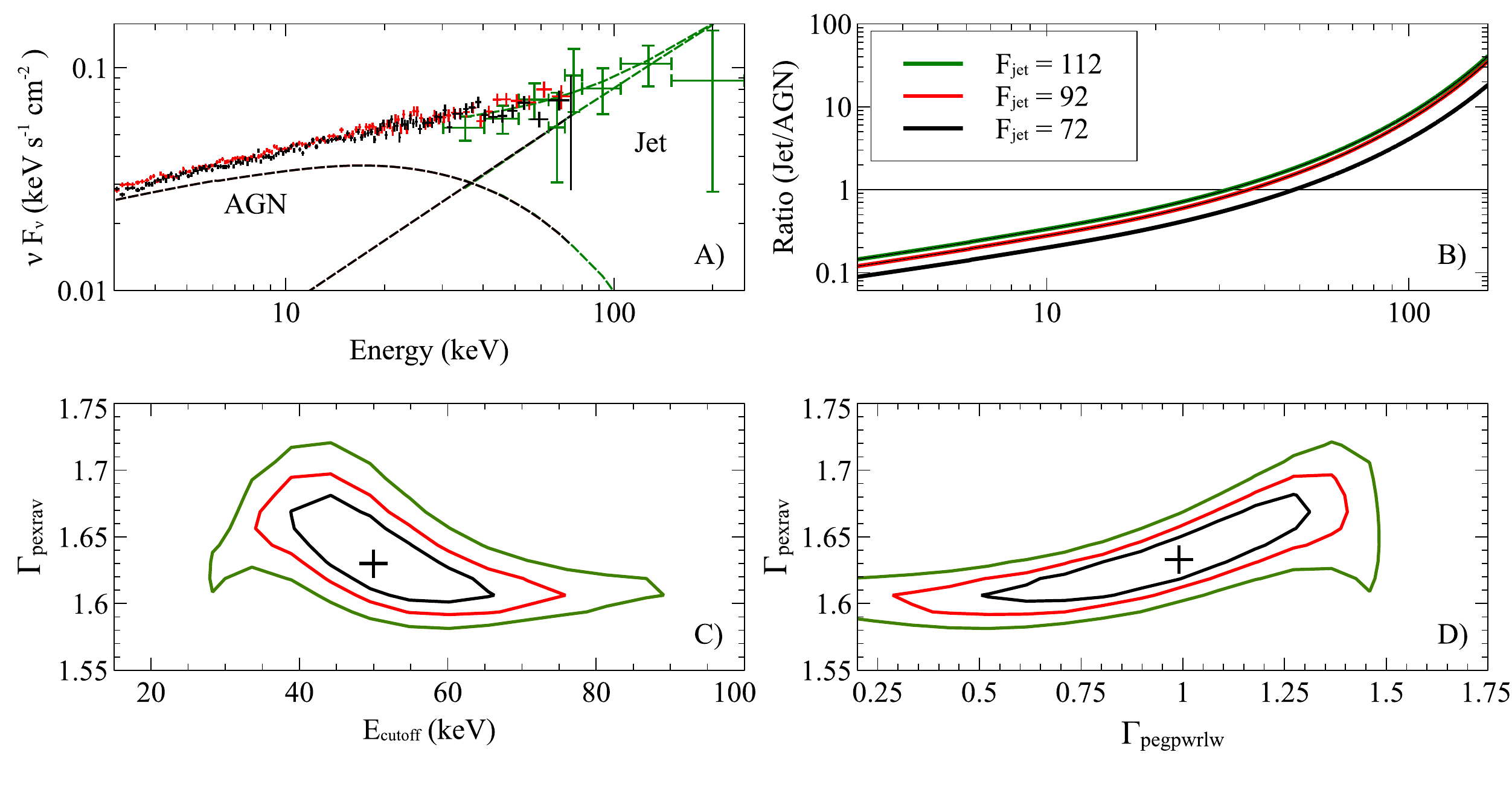}
\caption{\textit{INTEGRAL} and \textit{NuSTAR} fit. A) fit from Table \ref{tableNuIntegral} (black/red = FPMA/FPMB, green=ISGRI). B) Ratio between jet and AGN component for 3 flux settings of the jet (given in units of $10^{-12}$ erg cm$^{-2}$ s$^{-1}$). C) Contours of input spectrum of the \texttt{pexrav}. D) Contours of the photon index of the AGN against the jet component.}
\label{integralnustarfit}
\end{figure*}

Table \ref{tableNuIntegral} lists the best fit parameters. We explore the degeneracies between $\Gamma_\mathrm{pexrav}$, $E_\mathrm{cutoff}$, and $\Gamma_\mathrm{pegpwrlw}$ in Figure \ref{integralnustarfit}. The spectral energy distribution of the best fit with the two model components is shown in panel A.  Panel B shows the flux ratio between the AGN/jet for three different 80--150\,keV jet fluxes ($F_\mathrm{jet}$ = 72, 92, and 112$\times 10^{-12}$ erg cm$^{-2}$ s$^{-1}$). These fluxes span the $\pm 1\sigma$ level errors of the \textit{INTEGRAL} July fit (Table \ref{tableIntegral}), and for this range of fluxes the point where the jet crosses over the AGN component occurs between 30--50\,keV. At 6\,keV the AGN component is about an order of magnitude above the jet. Contours of $\Gamma_\mathrm{pexrav}$ versus $E_\mathrm{cutoff}$ are shown in panel C, and $\Gamma_\mathrm{pexrav}$ versus $\Gamma_\mathrm{pegpwrlw}$ in panel D. While these parameters are clearly coupled, they bound a well defined space, predicting the AGN photon index to be between 1.6--1.7 with a cutoff energy of 30--70 keV. The jet photon index lies somewhere between 0.5--1.5.

%We note here that 3C\,273 is a gamma-ray source detected by \textit{Fermi} \citep{Abdo2010}, but it showed relatively weak flux with no indication of flaring activity around the time of the \textit{NuSTAR} observation\footnote{http://fermi.gsfc.nasa.gov/ssc/data/access/lat/msl\_lc/}. While the \textit{Fermi} detection clearly supports the presence of the jet component, these observations could not be used to further constrain its spectral index.

\begin{figure}
\includegraphics[width=0.45\textwidth]{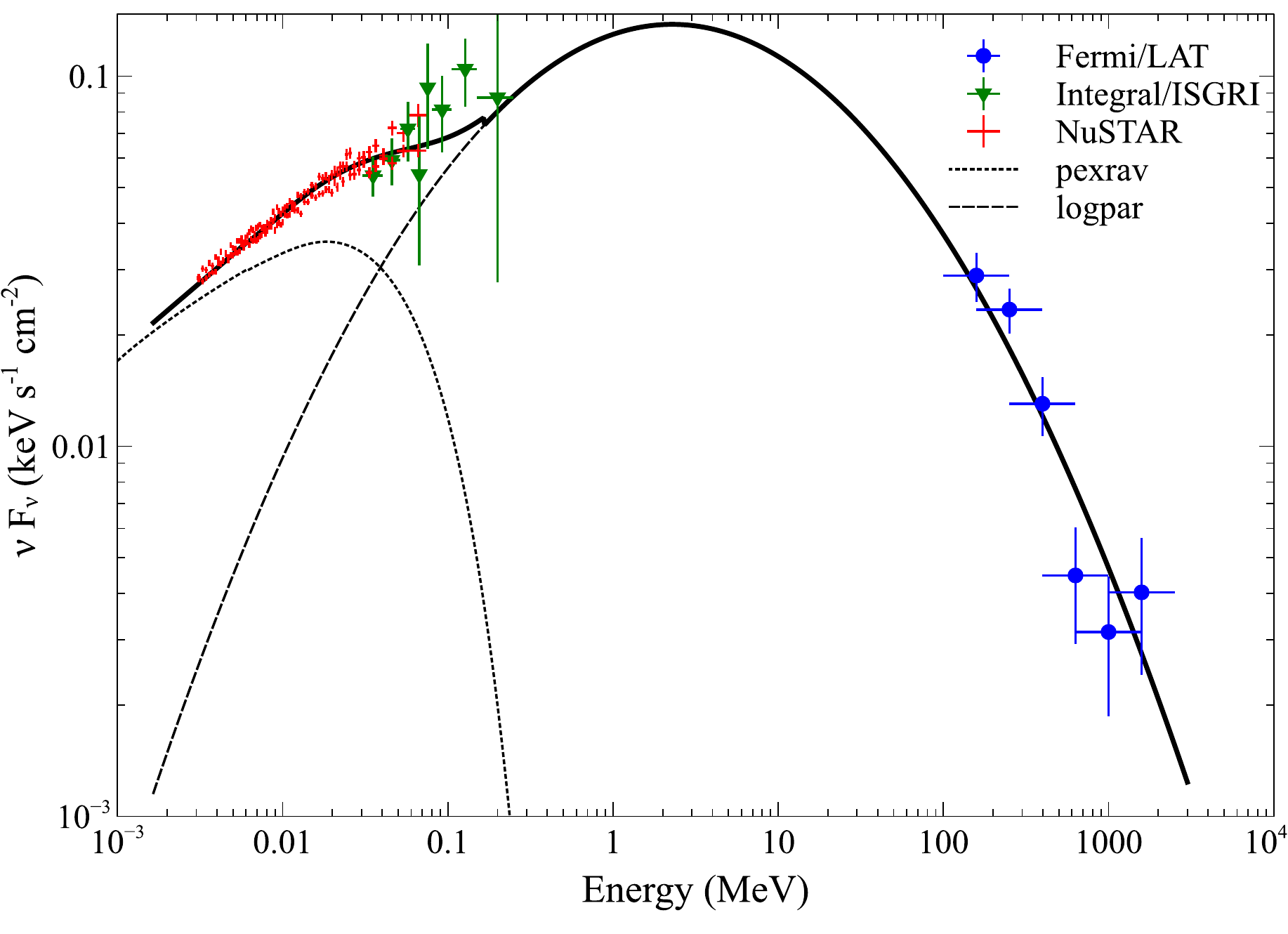}
\caption{Broad-band SED for 3C\,273 where the AGN component (\texttt{pexrav}) has been kept fixed to the values from Table \ref{tableNuIntegral} and the jet fitted with a log-parabola that peaks at $E_\mathrm{peak}=2.28 \pm 0.21$\,MeV.}
\label{fermi}
\end{figure}

To get a better constraint on the slope of the jet, we extracted a high-energy $\gamma$-ray ($>$ 100 MeV) spectrum of 3C\,273 from the \textit{Fermi}/LAT data. We assumed that the broad-band SED of the jet component is a log-parabola $N(E) = N_{\rm LogP}(E/E_0)^{[-\alpha-\beta\log(E/E_0)]}$, which has the advantage of adding only one parameter to the overall model. Parameter $\alpha$ corresponds to the local photon index at the photon energy of $E_0$, and here we choose $E_0 = 100\;{\rm keV}$. We added this model to the best-fit \texttt{pexrav} model from Table \ref{tableNuIntegral}, which we kept fixed, and we fitted it jointly to the \textit{NuSTAR}, \textit{INTEGRAL} and \textit{Fermi}/LAT data (Figure \ref{fermi}). We obtained a good fit with $\chi^2_{\rm red} = 1.117$, with the results $\alpha = 1.428 \pm 0.011$ and $\beta = 0.091 \pm 0.002$. This places the SED peak of the jet component at $E_{\rm peak} = (2.28 \pm 0.21)\;{\rm MeV}$, consistent with the results of \citet{Esposito2015}.

We note that the \integral\ data points are systematically above the fit. But a perfect fit is not expected, since 3C\,273 was found in a relatively low $\gamma$-ray state, which required integrating the $\gamma$-ray flux over a timescale of 50 days, far longer than the 6-day \nustar\ campaign. Short time scale variability might therefore have put the jet in a higher flux state during the \nustar\ period than on average. The fit, however, shows that the inferred slope of the jet is indeed reasonable.

\begin{table}
\centering
\caption{INTEGRAL Spectral fits: \texttt{pegpwrlw}}
\begin{tabular}{l|c|c|c|c}
\hline
Name & $\Gamma$ & Flux\footnote{Energy range 40--100\,keV in units of $10^{–12}$ erg cm$^{–2}$ s$^{-1}$.}& Flux\footnote{Energy range 80--150\,keV in units of $10^{–12}$ erg cm$^{–2}$ s$^{-1}$.} & $\chi^2_\mathrm{red}$/dof \\
\hline
December 2011 & $1.76 \pm 0.09$ & $118 \pm 5$ & $91 \pm 7 $ & 4.90/6 \\
June 2012 & $1.88 \pm 0.16$ & $128 \pm 10$ & $94 \pm 14 $ & 4.66/6 \\
July 2012 & $1.53 \pm 0.25$ & $105 \pm 13 $ & $92 \pm 20$ & 1.49/6 \\
\hline
\end{tabular}
\label{tableIntegral}
\end{table}

\begin{table}
\centering
\caption{INTEGRAL/NuSTAR June: \texttt{tbabs*const*(pexrav+pegpwrlw)}}
\begin{tabular}{l|c}
\hline
\multicolumn{2}{c}{\texttt{pexrav} (AGN)} \\
\hline
$\chi^2_\mathrm{ref}$ & 0.973 for 1478 dof \\
$\Gamma_{AGN}$ & 1.638 $\pm 0.045 $ \\
Cutoff Energy & 47 $\pm 15 $ keV\\
Relative reflection & 0.15 (frozen) \\
\hline
\multicolumn{2}{c}{\texttt{pegpwrlw} (jet)} \\
\hline
$\Gamma_{jet}$ & $1.05 \pm 0.4$ \\
Flux\footnote{$1\times 10^{-12}$ erg cm$^{-2}$ s$^{-1}$} (80--150\,keV) & 92 (frozen) \\
\hline
constant FPMA & 1.0 (frozen)\\
constant FPMB & 1.04 $\pm 0.01$\\
constant ISGRI & 0.95 $\pm 0.11$\\
\hline
\end{tabular}
\label{tableNuIntegral}
\end{table}

\subsection{The Iron Line}\label{ironcont}
%Since we do not have simultaneous coverage of \textit{NuSTAR} and \textit{INTEGRAL} across the entire \textit{NuSTAR} observation window, we turn now and in the subsequent sections to the \textit{NuSTAR} data set alone.

Although the detection of an iron line in 3C\,273 has in the past been firm, it remains faint and at times even absent. To maximize our signal we use all of the 244\,ks of the \textit{NuSTAR} observation and just \nustar\ alone. We restricted the energy range between 3--10\,keV. In this limited band an absorbed power-law provides a good fit, but with a slight excess between 5--7\,keV as can be observed in Figure \ref{cutoffratio}. We added a gaussian component with a fixed rest energy of 6.4\,keV and the quality of fit improves from $\chi^2/dof=419/345$ without to $\chi^2/dof=398/343$ with. A Monte-Carlo simulation shows this feature to be significant in excess of 99.97\%.

 The width of the line is broad, $\sigma_\mathrm{width} = 0.65 \pm 0.3$\,keV, and we measure an equivalent width of $EW = 23 \pm 11$\,eV, which is within the envelope of what has been previously reported \citep{Yaqoob2000,Kataoka2002,Page2004,Grandi2004}. 

We note that the magnitude of the iron emission is close to the calibration limit ($\pm 1\%$), and as a separate test we fixed the line energy to 6.4\,keV and allowed the redshift to vary. We recovered the correct redshift within errors. We also replaced the line with a relativistic \texttt{diskline} model but did not find significant improvement to the fit.

\subsection{Continuum Models}
As discussed in the previous sections, the 3C\,273 continuum data can be well fit with an absorbed cutoff power-law with weak reflection (a  \texttt{pexrav} model) up to the \nustar\ upper energy limit at 78\,keV. However, inclusion of the \textit{INTEGRAL} data reveals that the best fit between the two observatories requires an additional power-law for the jet component. Unfortunately, \textit{INTEGRAL} does not cover the entire \nustar\ observation, and the rise of the jet component is not visible in the \nustar\ data alone. We therefore proceed with our investigation of the continuum by using the full 244\,ks \nustar\ observation, but make the assumption that the jet component is constant for the entire time range and takes on the values given in Table \ref{tableNuIntegral}. We realize this to be an approximation and therefore to contrast we also fit the the data without the jet component to bracket parameters.

\begin{figure*}
\includegraphics[width=\textwidth]{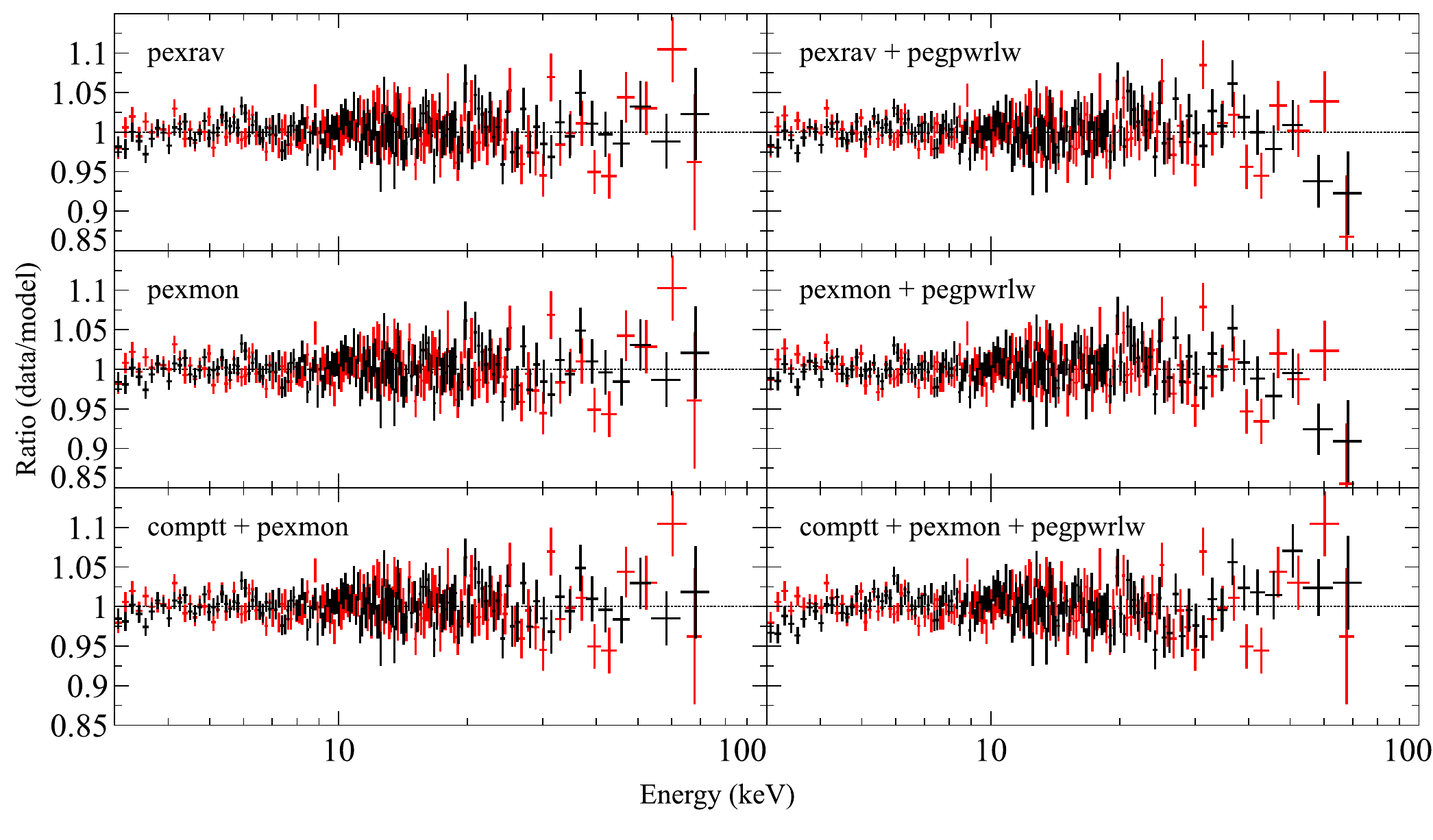}
\caption{Ratio plots of data to model for the fits in Table \ref{multimodels}. Not shown is the \texttt{pexrav+zgauss} since the improvement in these residuals is in the iron-line region covered in Figure \ref{cutoffratio}.}
\label{continum}
\end{figure*}

The results of the continuum fits are shown in Table \ref{multimodels} and the residual of data to model in Figure \ref{continum}. We first fit the continuum with a $\texttt{pexrav}$ model with and without a gaussian iron line included and compare with a \texttt{pexmon} \citep{Nandra2007}, which self-consistently includes the iron line. The two models yield very similar parameters with a photon index $\Gamma = 1.63 \pm 0.01$ and $E_\mathrm{cutoff} \sim 260 \pm 35$\,keV for the model without the jet, and $\Gamma = 1.66 \pm 0.01$ and $E_\mathrm{cutoff} \sim 50 \pm 2$\,keV with the jet. The relative reflection for both cases is around $R \sim 0.1$.

To probe a more physical model for the coronal continuum, we fitted with a \texttt{CompTT} model \citep{Titarchuk1994} and included the reflection component using \texttt{pexmon}. The \texttt{CompTT} model calculates the emergent spectrum resulting from Comptonization in an electron plasma, taking into account relativistic effects. It has models for a disk and spherical plasma geometry, and we adopted the disk geometry. We assumed solar abundances, fixed the inclination angle at 35$^\circ$, and used as soft photon temperature from the disk 10\,eV based on the temperature-luminosity relation from \citet{Makishima2000} for a 10\% accretion efficiency and $10^9$ M$_\odot$ black hole \citep{courvoisier1998}. We fixed the values of the reflection component to the values found for the \texttt{pexmon} fit, checking that the shape of the direct spectrum of the \texttt{pexmon} model matched that of the \texttt{CompTT} direct spectrum, and included only the reflection part of the model. 

Table \ref{multimodels} shows the results from fitting the \texttt{CompTT} disk model to the \nustar\ data with and without a jet component (with jet parameters again fixed to match the \integral\ $\mathrm{E}>80$~keV data as provided in Table~\ref{tableNuIntegral}). The inclusion of the jet component significantly changes the best fit parameters.  Without the jet, we find values of the plasma temperature of $kT_e = 247^{+69}_{-64}$\,keV, while with the jet included the temperature decreases to $kT_e = 12.0 \pm 0.3 \pm 1$\,keV and increases the optical depth from $\tau_p = 0.15_{-0.04}^{+0.08} $ without jet to $\tau_p = 2.77 \pm 0.06$ with the jet.  As shown (Figure \ref{continum}) the fit without jet is excellent and shows no indication of a jet component. The fit with jet displays residuals at high energies, indicating that our assumption of a constant jet is not correct, or that the constraints from \integral\ are not sufficient to properly constrain this component. Allowing the jet to fit freely we can find better solutions, but these results push the \texttt{CompTT} model up against its valid phase space and are thus questionable. We can, however, place an upper limit of $kT_e \sim 133^{+42}_{-29}$\,keV on the electron temperature, since inclusion of the jet in any form decreases the value from the no-jet scenario. Conversely the optical depth has a lower limit of $\tau_p = 0.33 \pm 0.1$. Given that the uncertainties in the \texttt{CompTT} physical parameters are completely driven by the uncertainties in the level of the jet flux, we do not investigate alternative coronal geometries further.

%A fixed jet model yields very tight constraints for the plasma electron temperature and optical depth. Allowing for a floating jet constrains the lower limit of the optical depth, but does not greatly affect the electron temperature. We find very reasonable values for the jet photon index, $\Gamma_\mathrm{jet} = 1.46 \pm 0.07$, and jet flux, $F(80-150)=83^{+8}_{-4} \times 10^{12}$ ergs\,s$^{-1}$\,cm$^{-2}$, with a plasma temperature for both geometries, $kT = 10 \pm 1$\,keV, and the optical depth high, $\tau_p(\mathrm{disk})=3.8^{+1.5}_{-1}$ and $\tau_p(\mathrm{sphere})=8.3^{+3.2}_{-1.9}$.

%We show the confidence contours and the best fit disk in Figure \ref{contourcomptt} between jet flux, $\tau_p$, and $kT_e$. The contours show that with decreasing electron temperature, the optical depth increases quite steeply. The ragged edge of $kT_e$ v. $\tau_p$ reflects the lower limit where the generalized Comptonization model in \texttt{CompTT} is considered valid according to \citet{Hua1995} (At 10\,keV. $\tau_p > 2$).  

We attempted to fit with a \texttt{CompPS} model \citep{Poutanen1996}, which is a coronal plasma model that includes reflection off a cold disk in the manner of the \texttt{pexrav}, but the returned fits were within a regime of high optical depth and low temperature where the model is unfortunately not considered reliable due to the large amount of scattering \citep{Poutanen1996}.

%We assumed solar abundances, fixed the inclination angle at 35$^\circ$ and assumed a multi-colored disk with inner temperature $T_{bb}$ = 10\,eV as we did for \texttt{CompTT}. Because the model does not calculate the iron line reflection we included a \texttt{zgauss} component and fixed the relative reflection to 0.15. As for \texttt{CompTT} we were able to leave the jet photon index unconstrained. We fitted both for the slab geometry and spherical geometry, but could only get the slab geometry to give a reasonable fit. We find values that are similar to those of the disk of \texttt{CompTT}, indicating a low temperature, $kT = 15.7 \pm 0.6$, and high optical depth, $\tau_p = 2.89 \pm 0.04$. 

%\textbf{In all cases where the iron line was not implicitly included in the model, we added a redshifted gaussian line (\texttt{zgauss} in XSPEC) to account for the iron line and found widths of around $EW \sim 20 \pm 10$ that were consistent with the presence of a faint, broad line.}

\begin{table}
\centering
\caption{\nustar\ continuum spectral fits.}
\begin{tabular}{l|c|c}
\hline
Parameter & w/o jet & w/ jet \\
\hline
\texttt{pexrav\footnote{Solar abundances.}} & \multicolumn{2}{c}{ (835 dof)} \\
\hline
$\chi^2_\mathrm{red}$ & 1.042 & 1.039 \\
$\Gamma$ & $1.63 \pm 0.01$ & $1.66 \pm 0.01$ \\
$E_\mathrm{cutoff}$ & $262 \pm 34$ keV & $52 \pm 2$ keV \\
Relative reflection & $0.07 \pm 0.03$ & $0.2 \pm 0.05$\\
\hline
\texttt{pexrav$^\mathrm{a}$+zgauss} & \multicolumn{2}{c}{ (833 dof)} \\
\hline
$\chi^2_\mathrm{red}$ & 1.033 & 1.035\\
$\Gamma$ & $1.63 \pm 0.01$ & $1.66 \pm 0.01$\\
$E_\mathrm{cutoff}$ & $260^{+38}_{-30}$ keV & $52 \pm 2$ keV \\
Relative reflection & $0.07 \pm 0.03$ & $0.2 \pm 0.03$\\
Line Energy (fixed) & 6.4 keV & 6.4 keV \\
Line width $\sigma$ & $0.62^{+0.50}_{-0.42}$ keV & $0.87^{+2.6}_{-0.7}$ keV\\
EW & $15 \pm 11$ eV & $15 \pm 11$ eV\\
\hline
\texttt{pexmon$^\mathrm{a}$} & \multicolumn{2}{c}{ (835 dof)} \\
%\multicolumn{2}{c}{incl = 85, Abund = 1, Fe Abund = 1} \\
\hline
$\chi^2_\mathrm{red}$ & 1.035 & 1.055\\
$\Gamma$ & $1.63 \pm 0.01$ & $1.64 \pm 0.01$ \\
$E_\mathrm{cutoff}$ & $257^{+37}_{-28}$\,keV & $53 \pm 2$\,keV\\
Relative Reflection & $0.04 \pm 0.01$ & $0.08 \pm 0.02$\\
\hline
\texttt{CompTT$^\mathrm{a}$+pexmon} & \multicolumn{2}{c}{ (840 dof)} \\
\hline
$\chi^2_\mathrm{red}$ & 1.040 & 1.145 \\
Geometry & disk & disk \\
T$_0$ (fixed) & 10\,eV & 10\,eV\\
Plasma temp, kT & $133^{+42}_{-29}$ keV & $12.0 \pm 0.3$ keV\\
$\tau_p$ & $0.33_{-0.09}^{+0.11}$ & $2.77 \pm 0.06 $ \\
$\Gamma$ (fixed) & 1.63 & 1.64 \\
$E_\mathrm{cutoff}$ (fixed) & 257\,keV & 53\,keV\\
Relative Reflection (fixed) & 0.04 & 0.08\\
\hline
\end{tabular}
\label{multimodels}
\end{table}

\subsection{Variability}\label{variable}
\begin{figure}
\includegraphics[width=0.45\textwidth]{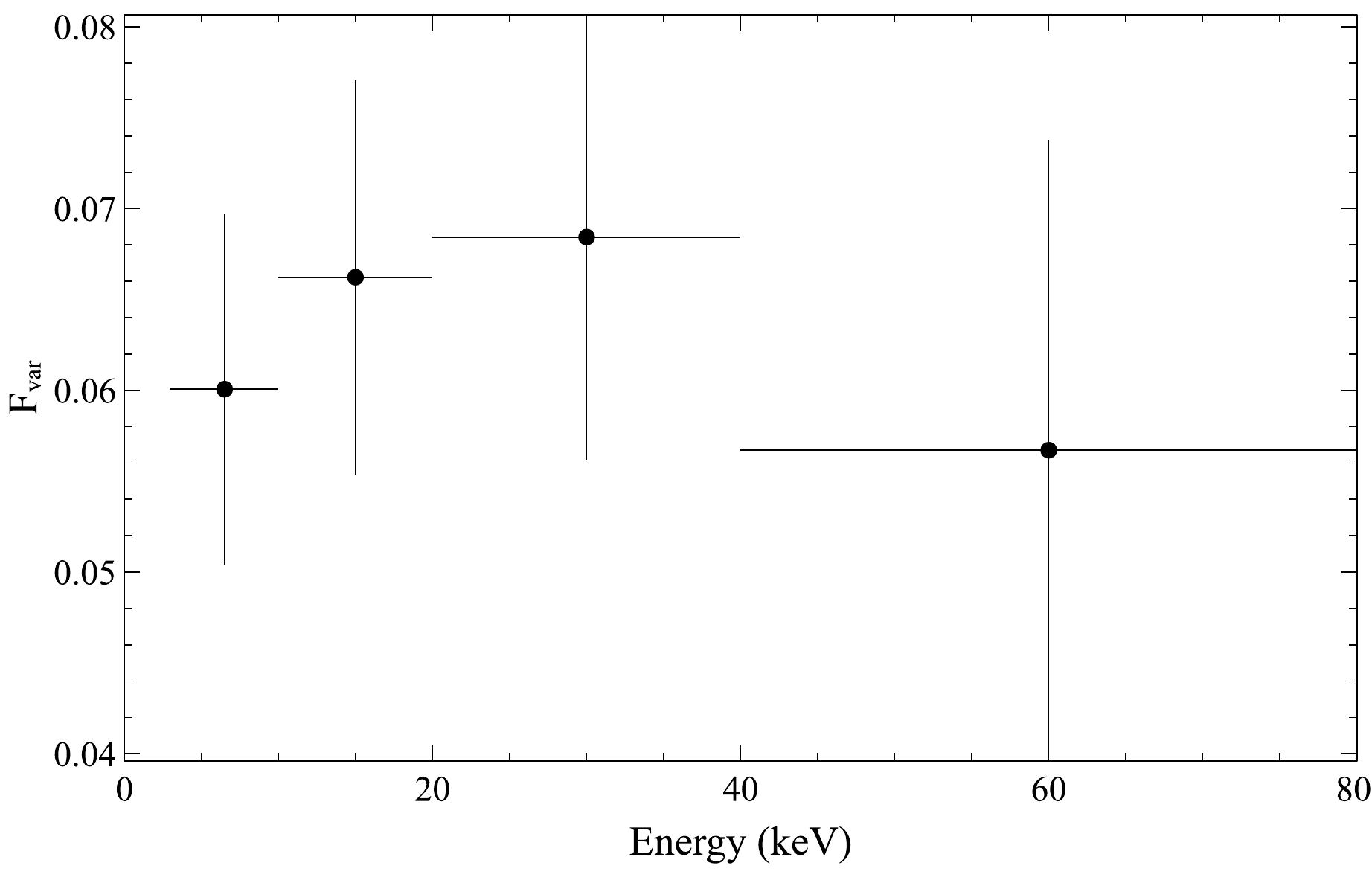}
\caption{Noise subtracted variance fraction of the mean for the four energy bands: 3--10, 10--20, 20--40, and 40--78\,keV. Statistically there is no evidence of a difference in variability as a function of energy.}
\label{fvar}
\end{figure}

The \textit{NuSTAR} lightcurve (Figure \ref{lightcurve}) shows that 3C\,273 went through flux changes of 10 -- 30\% during the observation. We calculated the noise subtracted variance fraction of the mean, $F_\mathrm{var}$, using the formalism of \citet{Edelson2002}, and Figure \ref{fvar} shows $F_\mathrm{var}$ statistically does not vary as a function of energy during our observation.

\begin{figure}
\includegraphics[width=0.45\textwidth]{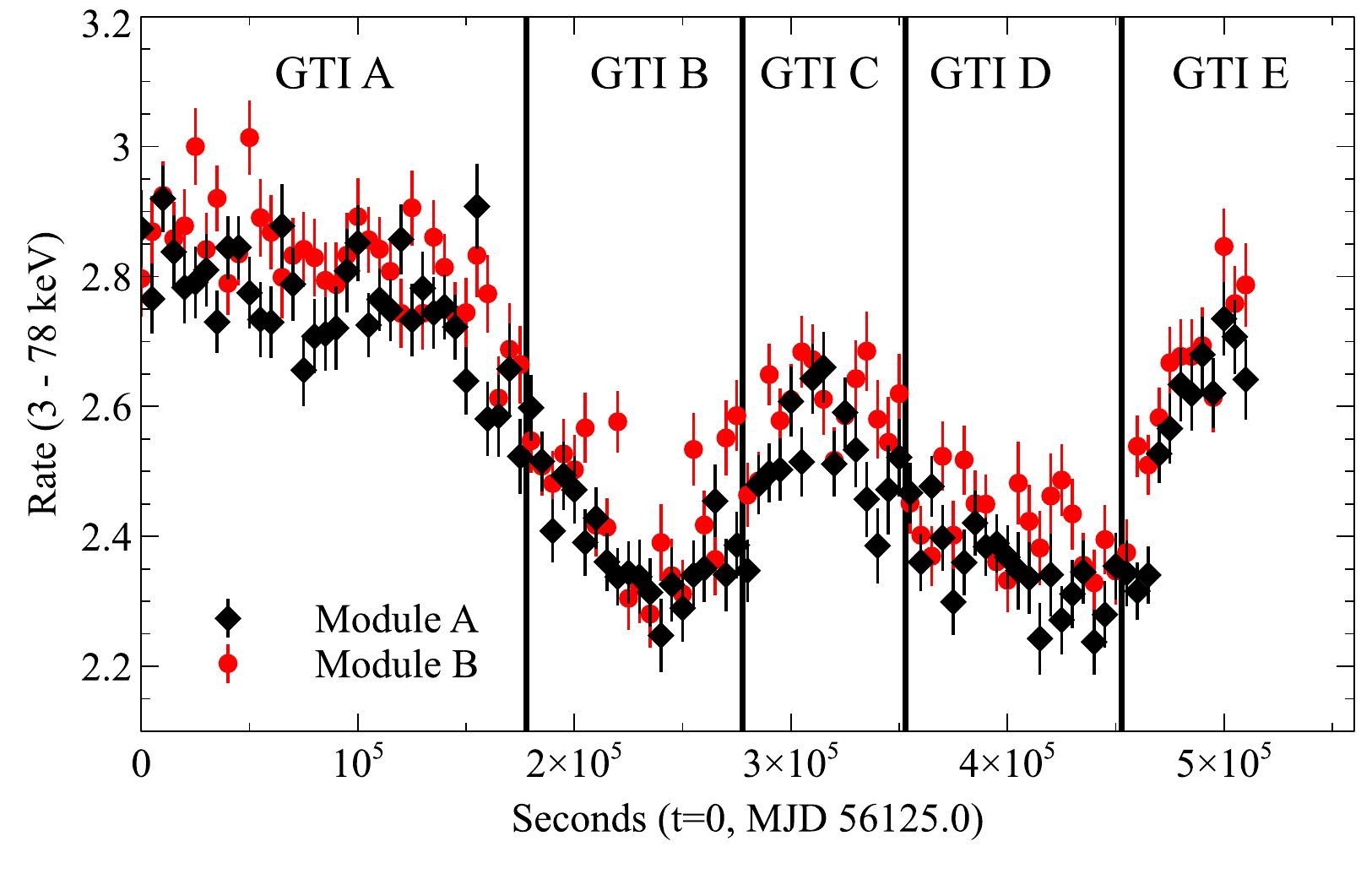}
\caption{Light-curve showing GTI splits picked for periods where the flux remained roughly constant.}
\label{lightcurvewithgti}
\end{figure}

We investigated the variability of the spectrum by decomposing the observation into 5 time bins, GTI\,A-GTI\,E (start and stop times are listed in Table \ref{tablevariable}), as illustrated in Figure \ref{lightcurvewithgti}. The motivation behind this partitioning is to cover the sections where the flux is roughly constant. We chose to continue with the \texttt{pexrav} model, and fitted both with and without the jet component, assuming it was frozen to the values given in Table \ref{tableNuIntegral}. The two models and the parameters are summarized in Table \ref{tablevariable}.

We find that by enforcing a jet component we obtain lower $E_\mathrm{cutoff}$ energy since more bending is required when the spectrum is diluted by the jet, and the jet component systematically softens the index. The jet component also systematically increases the relative reflection for the same reason.

\begin{figure}
\includegraphics[width=0.45\textwidth]{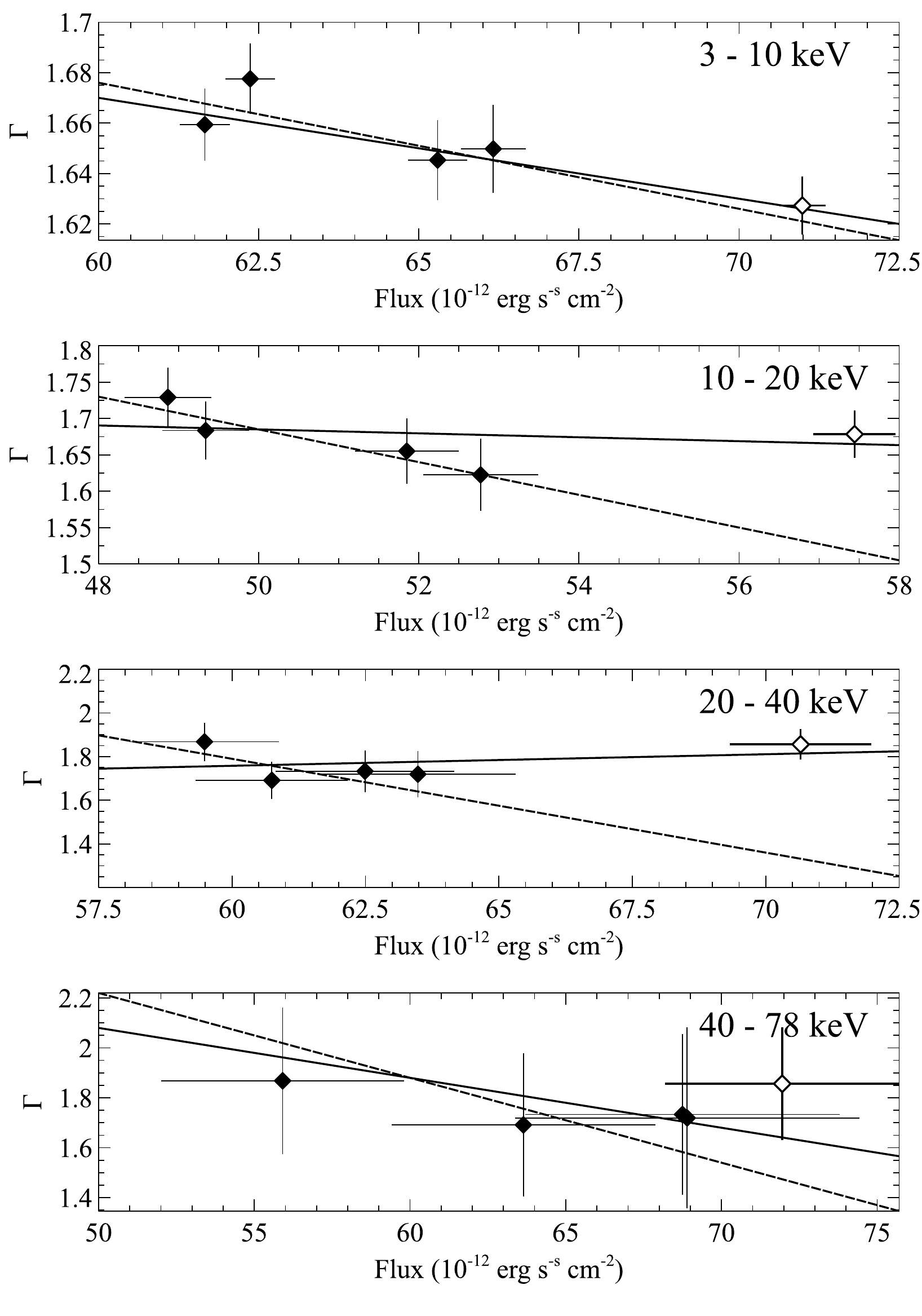}
\caption{Flux in the respective bands 3--10, 10--20, 20--40, and 40--78\,keV as a function of a phenomenological power-law index fitted for each interval, GTI\,A--E, shown in Figure \ref{lightcurvewithgti}. The unfilled points correspond to the interval GTI\,A. The solid line is a fit to all points, and the dashed line for intervals GTI\,B--E alone.}
\label{fluxvar}
\end{figure}

Because of the coupling between the photon index, cutoff energy, and relative reflection, it is hard to determine the correlation between flux and slope for the AGN component. To approach this question in a model independent manner, we decomposed the spectrum into four energy bands, 3--10, 10--20, 20--40, and 40--78\,keV, and fit each section with a phenomenological power-law, calculating the flux for the intervals. The results are shown in Figure \ref{fluxvar}. The highest flux points marked with an open symbol is GTI\,A. We fitted a linear function to all data points (solid line) and to GTI\,B-GTI\,E excluding the GTI\,A point (dashed line). 

The linear Pearson correlation \citep{Press1992} favors in all but the lowest energy range the GTI\,B-GTI\,E (P1 series) over GTI\,A-GTI\,E (P2 series): (3--10\,keV): P1/P2 = -0.73/-0.86, (10--20\,keV): P1/P2 = -0.79/-0.64, (20--40\,keV):  P1/P2 = -0.64/0.14, and (40--78\,keV):  P1/P2 = -0.70/-0.62. In the lowest energy band we find that the photon index is inversely correlated with the flux, becoming harder for higher fluxes. This correlation persists through all bands when excluding GTI\,A, but starts to break down in the 10--20\,keV band when GTI\,A is included. Since the source during GTI\,A was very stable compared to later times, it might suggest the source was in a different state altogether.

%\begin{figure*}
%\includegraphics[width=\textwidth]{plots/contour_comptt.pdf}
%\caption{Confidence Contour plots for the \texttt{CompTT} best fit for a disk. The confidence contours for the spherical have %the same shape, but are shifted in $\tau_p$ at the best fit value of $\tau_p=8.3$. The contours show that the optical depth %quickly increases for low electron temperatures.}
%\label{contourcomptt}
%\end{figure*}

%\begin{figure*}
%\includegraphics[width=\textwidth]{comptt_pex_resid.pdf}
%\caption{\textbf{Left panel}: Confidence contour plots of $kT_e$ and $\tau_p$ for disk geometry with jet. \textbf{Right panel}: Ratio %of data to model of the best fit without and with a fixed jet component. The high energy residuals indicate the assumption of a %constant jet is likely not accurate.}
%\label{contourcomptt}
%\end{figure*}

\begin{table*}
\centering
\caption{NuSTAR Spectral fits: \texttt{tbabs $\times$ pexrav}}
\begin{tabular}{l|c|c|c|c|c|c}
\hline
Name & START & STOP & $\Gamma$ & $E_\mathrm{cutoff}$ (keV) & Relative reflection & $\chi^2_\mathrm{red}$/dof \\
\hline
\multicolumn{7}{c}{w/o \texttt{jet}} \\
\hline
%GTI 1 & 2012:196:00:36:14 & 2012:197:11:29:47 & $1.60 \pm 0.02$ & $187_{-27}^{+38}$ & $0.09 \pm 0.05$ & 1.002 1209\\
%GTI 2 & 2012:197:11:29:47 & 2012:198:15:53:19 & $1.64 \pm 0.02$ & $300_{-78}^{+164}$ & $0.03^{+0.06}_{-0.03} $ & 1.028/1090\\
%GTI 3 & 2012:198:15:53:19 & 2012:199:12:43:19 & $1.67 \pm 0.03$ & $292_{-81}^{+176}$ & $0.14 \pm 0.08 $ & 0.965/930\\
%GTI 4 & 2012:199:12:43:19 & 2012:200:23:11:58 & $1.64 \pm 0.02$ & $316_{-76}^{+146}$ & $0.06 \pm 0.06$ & 1.052/1133\\
%GTI 5 & 2012:200:23:11:58 & 2012:201:23:28:43 & $1.63 \pm 0.02$ & $299_{-80}^{+168}$ & $0.04^{+0.06}_{-0.04}$ & 1.052/1003\\
%\hline
GTI\,A & 2012:196:00:36:14 & 2012:198:01:59:59 & $1.61 \pm 0.02$ & $213_{-32}^{+46}$ & $0.07 \pm 0.04$ & 1.056/1360\\
GTI\,B & 2012:198:01:59:59 & 2012:199:05:46:39 & $1.67 \pm 0.02$ & $257_{-59}^{+105}$ & $0.13 \pm 0.07$ & 0.993/1036\\
GTI\,C & 2012:199:05:46:39 & 2012:200:02:36:39 & $1.63 \pm 0.03$ & $265_{-70}^{+141}$ & $0.05^{+0.07}_{-0.05}$ & 1.025/1141\\
GTI\,D & 2012:200:02:36:39 & 2012:201:06:23:19 & $1.65 \pm 0.02$ & $485_{-175}^{+622}$ & $0.05^{+0.06}_{-0.05}$ & 0.978/1035\\
GTI\,E & 2012:201:06:23:19 & 2012:201:23:28:43 & $1.63 \pm 0.03$ & $293_{-85}^{+203}$ & $0.07^{+0.08}_{-0.07}$ & 1.047/1006\\
\hline
\multicolumn{7}{c}{w/ \texttt{jet}} \\
%\hline
%GTI 1 & 2012:196:00:36:14 & 2012:197:11:29:47 & $1.63 \pm 0.03$ & $53 \pm 4$ & $0.23 \pm 0.09$ & 1.001/1205 \\
%GTI 2 & 2012:197:11:29:47 & 2012:198:15:53:19 & $1.66 \pm 0.03$ & $53 \pm 6$ & $0.13 \pm 0.1$ & 1.097/1088 \\
%GTI 3 & 2012:198:15:53:19 & 2012:199:12:43:19 & $1.71 \pm 0.04$ & $49 \pm 6$ & $0.34 \pm 0.15 $ & 1.002/928 \\
%GTI 4 & 2012:199:12:43:19 & 2012:200:23:11:58 & $1.67 \pm 0.03$ & $52 \pm 5$ & $0.20 \pm 0.10$ & 1.051/1130 \\
%GTI 5 & 2012:200:23:11:58 & 2012:201:23:28:43 & $1.65 \pm 0.02$ & $53 \pm 6$ & $0.14 \pm 0.11$ & 1.054/1001 \\
\hline
GTI\,A & 2012:196:00:36:14 & 2012:198:01:59:59 & $1.64 \pm 0.02$ & $54 \pm 4$ & $0.19 \pm 0.07$ & 1.051/1360 \\
GTI\,B & 2012:198:01:59:59 & 2012:199:05:46:39 & $1.69 \pm 0.03$ & $45 \pm 5$ & $0.33 \pm 0.13$ & 0.996/1041 \\
GTI\,C & 2012:199:05:46:39 & 2012:200:02:36:39 & $1.65 \pm 0.04$ & $48 \pm 6$ & $0.20 \pm 0.13$ & 1.038/956 \\
GTI\,D & 2012:200:02:36:39 & 2012:201:06:23:19 & $1.68 \pm 0.03$ & $55 \pm 7$ & $0.16 \pm 0.11$ & 0.977/1035 \\
GTI\,E & 2012:201:06:23:19 & 2012:201:23:28:43 & $1.66 \pm 0.04$ & $55 \pm 8$ & $0.18 \pm 0.13$ & 1.039/893 \\
\hline
\end{tabular}
\label{tablevariable}
\end{table*}

\section{Discussion}
From the radio to $\gamma$-ray bands the SED of 3C\,273 is dominated by non-thermal radiation from the jet, however, evidence for emission from the inner accretion flow has been argued based on the presence of a weak iron line and soft excess in the X-ray, and the big blue-bump in the optical/UV, which are all common features of AGN \citep{Kataoka2002, Grandi2004, Pietrini2008}. Evidence for hard X-ray emission from the corona in the form of a cutoff power-law and reflection features above 10~keV have been elusive, perhaps due to the available instrument sensitivities or because the source was observed in states with high jet flux, which could have obscured the AGN signatures. 

%The \textit{NuSTAR} data show that the 3--78 keV spectrum cannot be fit simply by an absorbed power-law, but the spectrum requires a roll over above $\sim$20\,keV. Modeling the \nustar\ data alone we find a good fit using a cutoff power-law, and we find some evidence for a weak reflected continuum consistent with reflection from a cold accretion disk or distant material. The cutoff in the power-law at $262 \pm 34$~keV indicates this continuum originates from disk radiation Compton up-scattered in a hot coronal electron plasma of finite temperature. We confirm the presence of a weak iron line.  Assuming it is neutral, we measure its parameters to be $\sigma = 0.65\pm0.3$\,keV with an EW = $23\pm11$\,eV,  consistent with previous measurements \citep{Yaqoob2000,Kataoka2002,Page2004,Grandi2004}. If we include a reflection component we can account for the weak line and excess above 10~keV (the compton hump) with a reflection fraction R$= 0.07 \pm 0.03$.

The \textit{NuSTAR} data show that the 3--78 keV spectrum cannot be fit simply by an absorbed power-law, but the spectrum requires a roll over above $\sim$20\,keV. Modeling the \nustar\ data alone we find evidence for a weak reflected continuum consistent with reflection from a cold accretion disk or distant material. We confirm the presence of a weak iron line and assuming it is neutral, we measure its parameters to be $\sigma = 0.65\pm0.3$\,keV with an EW = $23\pm11$\,eV, consistent with previous measurements \citep{Yaqoob2000,Kataoka2002,Page2004,Grandi2004}. If we include a reflection component we can account for the weak line and excess above 10~keV (the compton hump) with a reflection fraction R$= 0.07 \pm 0.03$ and E$_\mathrm{cutoff}$=$262 \pm 34$~keV.

Considering \nustar\ together with \textit{INTEGRAL} data for a temporally overlapping window we show that the broadband (3--150~keV) spectrum cannot be explained with a single spectral model. Extrapolating the \nustar\ best-fit AGN model above 80~keV reveals a clear excess in \integral\ that we associate with emission from the jet. If we attempt to fit the combined \integral\ and \nustar\ spectrum with a model including the AGN components and an additional power-law to describe the jet emission, we find strong degeneracies and no unique fit. We therefore fix the flux of the jet component from \textit{INTEGRAL} 80--150~keV data and investigate the range of fits to the broadband spectrum corresponding to the range of jet flux uncertainty. To further limit degeneracies we fix the relative reflection to R$ = 0.15$. We find that the photon index for the coronal component ranges from 1.6 - 1.7, with a cutoff energy of 30--70~keV. The jet photon index is poorly constrained, ranging from 0.5--1.5. 

Taking into account the high-energy $\gamma$-ray spectrum from the \textit{Fermi}/LAT, and assuming a log-parabolic shape of the jet component, we find the local photon index at 100\,keV of about 1.4. We caution that the complete lack of data in the 1--100\,MeV range means that the broad-band shape of the jet component is uncertain. Nevertheless, the log-parabola model predicts a modest luminosity of the SED peak at 2\,MeV. If the photon index at 100\,keV was in the range 0.5--1, the implied peak luminosity of the jet component would need to be significantly higher, and that would be more demanding from the point of view of jet energetics. We note that a similar log-parabola model added to an AGN component was found to be successful by \citet{Esposito2015}, who analyzed the \textit{Swift}, \textit{INTEGRAL} and \textit{Fermi} data for 3C\,273 in different spectral states. In any case, we find that the jet component begins to dominate over the AGN component in the range from 30--50~keV. 

The cutoff energy range of 50--70\,keV found with the jet included, is at the low end compared to other sources where a cutoff energy has been measured by \textit{NuSTAR} \citep{Ballantyne2014,Brenneman2014,Marinucci2014,Matt2015,Balokovic2015}, however we note that considerable uncertainty arises due to the poor constraints on the jet emission. If we exclude the jet component an upper limit for the energy cutoff is $E_\mathrm{cutoff} \sim 260 \pm 40$\,keV (see Table \ref{multimodels}).

We explore a physical model for the AGN component and fit with the coronal \texttt{CompTT}, which models spectra from a Comptonizing coronal electron plasma \citep{Titarchuk1994}. If we fix the jet component parameters at the best-fit values given in Table~\ref{tableNuIntegral}, we find an unusually low plasma temperature of $kT = 12 \pm 1$\,keV, and high optical depth $\tau_p=2.77 \pm 0.06$. Without the jet component included we find $kT = 133^{+42}_{-29}$\,keV, and  optical depth $\tau_p=0.33 \pm 0.1$. Since the jet contribution is poorly constrained we can only conclude that the actual values lie between these extremes.

The reflection fractions we derive for all models we consider are low (R$ = 0.02 - 0.2$)  (See Table \ref{multimodels} and \ref{tablevariable}). These indicate very weak reflection, as is often found in broad-line radio galaxies (BLRGs) \citep{Wozniak1998,Eracleous2000,Zdziarski2001,Sambruna2009,Ballantyne2014}. Possible explanations for the weak reflection include high inner disk ionizations \citep{Ballantyne2002}, a change in inner disk geometry \citep{Eracleous2000,Lohfink2013}, obscuration of the the central accretion flow by the jet \citep{Sambruna2009}, black holes with retrograde spin \citep{Evans2010}, and dilution of the X-ray spectrum by jet emission \citep{Grandi2002}. Since we still only find a small reflection fraction even when taking into account the jet, it is likely not due to dilution of the jet. It has instead been suggested that a geometrically thick accretion disk, which obscures part of itself from the external source \citep{Paltani1998} could be responsible for the low reflection fraction. Considering the possibility of a high optical thickness of the corona, it could also be that the corona itself is smearing out the reflection signature.

3C\,273 is well known for its variability. During the \textit{NuSTAR} observation it went through several flux changes. Our study found no statistical evidence for a change in variability as a function of energy in the \textit{NuSTAR} band. The disentanglement and correlation of the jet and AGN component with respect to flux remains unclear. In the low energy band (3--10\,keV) we find an inverse correlation between flux and slope, hardening the spectrum with increasing flux. This correlation seems to persist through all bands when excluding the highest flux bin, but appears to break down above 10\,keV band when including the high flux bin. This inverse correlation in the low-energy band has been previously observed in the 2--10\,keV flux band by \citet{Kataoka2002} during their 1999--2000 campaign with \textit{RXTE}. However, during their 1996-97 campaign this correlation was no longer present. They attributed this change to the emergence of the AGN component during 1999--2000 when the source was at a higher flux than in 1996-97. We postulate that it may rather be that a higher flux level of the jet is responsible for the hardening of the spectrum. An inverse correlation is not common for AGN, where spectra typically get steeper with increasing flux (but we note that the opposite is generally true for blazars, both those associated with quasars, and the lineless variety;  respective examples are 
3C279 \citep{Hayashida2015}, and Mkn 421 \citep{Balokovic2013}). This could indicate that the variability is instead driven by the jet flux, hardening the spectrum as the jet flux contributes more strongly. The highest flux bin corresponding to GTI\,A does break the correlation, but considering the change in the light curve itself, there is an indication the source transitioned from one state to another at the end of GTI\,A. 

If we assume the two component scenario, it appears the AGN was bright at the beginning, clearly outshining the jet. Subsequently, lower flux softened the index at all energies, but it is not possible to tell what combination of jet and AGN flux levels went into generating the variability since the two components are not expected to be correlated \citep{Soldi2008}.

A long standing issue in 3C\,273 is that the Comptonization of UV photons predicts a correlation to exist between UV and X-ray flux, which was claimed to have been observed in the past by \citet{Walter1992} but later found absent by \citet{Chernyakova2007}. In the jet/AGN component scenario the superposition of two uncorrelated model components would affect the reliability of a correlation between UV and X-rays, and the disappearance of it may simply therefore be a matter of jet domination.

The spectral structure inferred by the consideration of the \textit{NuSTAR} and \textit{INTEGRAL} data could possibly also be explained by invoking a two-component inverse Compton model arising in a relativistic jet. In such a scenario, the MeV-to-GeV range emission is likely to be dominated by scattering of the isotropic radiation external to the jet \citep[as in External Radiation Compton, or ERC;  see, e.g.][]{Sikora1994}, while the slight ``hump'' detected by \textit{NuSTAR} and modeled by us as Compton reflection from the accretion disk would be due to synchrotron self-Compton (SSC) emission, with the target photons being the synchrotron photons in the jet itself \citep{Maraschi1992}. We consider such a scenario less likely: the \textit{NuSTAR} data (as well as previous X-ray observations) detect a broad Fe K line, which is a signature of reprocessing from the accretion disk, which is likely to be accompanied by the reflection component as in our modeling. Finally, as argued by \citet{Esposito2015}, the timing analysis performed by those authors argues for the presence of the Seyfert component in addition to the jet component.  However, any firm conclusions regarding the origin of the jet radiation - namely the relative dominance of the SSC vs. ERC component - are not possible with the data presented here.

\section{Conclusion}
In the coordinated observing campaign in 2012 we observed 3C\,273 in a state where the jet flux was relatively weak. This made it possible to observe spectral signatures from coronal emission and constrain the weak Compton reflection in the hard X-ray band.   We can separate the spectral components from the inner accretion flow from the jet flux in the 3 - 150~keV band quantitatively only by making an explicit assumption about jet flux level. To truly separate the two components and constrain the physical parameters of the corona, and/or address the possibility of SSC emission from the jet masking as reflection, further simultaneous observations with \nustar, \integral\ and \textit{Fermi} of the source in a state with low levels of flux from the jet are required.

{\it Facilities:} \facility{Fermi}, \facility{CXO}, \facility{INTEGRAL}, \facility{NuSTAR}, \facility{Suzaku}, \facility{Swift}, and \facility{XMM}

\acknowledgments
We would like to thank Chris Done for bringing to our attention the alternative interpretation of the jet as two-component inverse Compton model, and the anonymous referee whose remarks and corrections helped improve the quality of this paper. This work was supported under NASA Contract No. NNG08FD60C, and made use of data from the \textit{NuSTAR} mission, a project led by the California Institute of Technology, managed by the Jet Propulsion Laboratory, and funded by the
National Aeronautics and Space Administration. We thank the \textit{NuSTAR} Operations, Software and Calibration teams for
support with the execution and analysis of these observations. This research has made use of the \textit{NuSTAR} Data Analysis
Software (NuSTARDAS) jointly developed by the ASI Science Data Center (ASDC, Italy) and the California Institute of Technology (USA).

\bibliography{bib}
\bibliographystyle{jwapjbib}

\end{document}